\documentclass[conference]{IEEEtran}
\IEEEoverridecommandlockouts
\usepackage{cite}
\usepackage{amsmath,amssymb,amsfonts}
\usepackage{algorithmic}
\usepackage{graphicx}
\usepackage{textcomp}
\usepackage{xcolor}

\usepackage{cite}
\usepackage{amsmath,amssymb,amsfonts}
\usepackage{algorithmic}
\usepackage{graphicx}
\usepackage{textcomp}
\usepackage{xcolor}
\usepackage{graphicx}
\usepackage{multirow}
\usepackage[normalem]{ulem}
\useunder{\uline}{\ul}{}
\usepackage{caption}
\usepackage{subcaption}
\usepackage{csquotes}
\usepackage{cleveref}
\usepackage{float}
\usepackage[hyphens]{url}
\usepackage{csquotes}

\newcommand{\james}[1]{\textcolor{black}{#1}}
\newcommand{\ehsan}[1]{{\textcolor{black}{#1}}}

\def\BibTeX{{\rm B\kern-.05em{\sc i\kern-.025em b}\kern-.08em
    T\kern-.1667em\lower.7ex\hbox{E}\kern-.125emX}}
\begin{document}

\title{A Reddit Dataset for the Russo-Ukrainian Conflict in 2022}

\author{
\IEEEauthorblockN{
Yiming Zhu\IEEEauthorrefmark{1}, 
Ehsan-Ul Haq\IEEEauthorrefmark{1}, 
Lik-Hang Lee\IEEEauthorrefmark{2}, 
Gareth Tyson\IEEEauthorrefmark{1}, 
% TBD\IEEEauthorrefmark{1}, 
and Pan Hui\IEEEauthorrefmark{1}\IEEEauthorrefmark{3}}
\IEEEauthorblockA{\IEEEauthorrefmark{1}Hong Kong University of Science and Technology}
\IEEEauthorblockA{\IEEEauthorrefmark{2}Korea Advanced Institute of Science and Technology}
\IEEEauthorblockA{\IEEEauthorrefmark{3}University of Helsinki, Helsinki}
Email: \{yzhucd, euhaq\}@connect.ust.hk  \quad likhang.lee@kaist.ac.kr
 \quad \{gtyson, panhui\}@ust.hk
}

\maketitle

\begin{abstract}
Reddit consists of sub-communities that cover a focused topic. This paper provides a list of relevant subreddits for the ongoing Russo-Ukrainian crisis. We perform an exhaustive subreddit exploration using keyword search and shortlist 12 subreddits as potential candidates that contain nominal discourse related to the crisis. These subreddits contain over 300,000 posts and 8 million comments collectively. We provide an additional categorization of content into two categories, \enquote{R-U Conflict}, and \enquote{Military Related}, based on their primary focus. We further perform content characterization of those subreddits. The results show a surge of posts and comments soon after Russia launched the invasion. \enquote{Military Related} posts are more likely to receive more replies than \james{\enquote{R-U Conflict} posts}.
% \gareth{Think one of the enquotes is wrong :) I assume one is supposed to be R-U Conflict?}
Our textual analysis shows an apparent preference for the Pro-Ukraine stance in \enquote{R-U Conflict}, while \enquote{Military Related} retain a neutral stance.
\end{abstract}


\section{Introduction}
On 24 February 2022, Russia launched what they refer to as a \emph{special military operation} into Eastern Ukraine after their announcement to recognize the Donetsk and Luhansk people's republics.\footnote{\url{https://edition.cnn.com/2022/02/23/europe/russia-ukraine-putin-military-operation-donbas-intl-hnk/index.html}} This led to the advancement of Russian military forces, attracting global attention.

Various organizations and individuals worldwide have been sharing their opinions on the crisis. 
Frequently, these expressions of diversified opinion carry support for either of the parties. However, such expressions are sometimes put forward with the objective of opinion engineering~\cite{haq2022weaponising}. Social media and online discussion forums can help researchers to explore essential research questions, including general opinion and sentiment on the topic, identifying and counteracting fake news, propaganda campaigns, as well as population movements.~\cite{kouloumpis2011twitter,shu2017fake,haq2020survey}

Such studies rely on social media datasets as a fundamental part of observing the questions mentioned above. Previously, researchers have collected and shared such datasets related to various social events, such as COVID-19 or Elections~\cite{zarei2020first,allcott2017social}. There are also already several datasets for the Russo-Ukrainian crisis from different social media, such as Twitter~\cite{haq2022twitter}, and Weibo~\cite{fung2022weibo}. Here, we present the first dataset from Reddit, a globally used discussion forum. We aim to present the key subreddits relevant to the crisis and describe the data contained within.

Unlike Twitter and Facebook, where people mostly submit short texts, Reddit users can submit longer posts containing more content. Furthermore, Reddit is themed around topic-specific communities called subreddits. Users can interact with posts by commenting and voting, making Reddit an ideal source to explore and study social information about crisis events \cite{bunting2021socially,wasike2011framing}. Therefore, to facilitate further related analysis, we present an open Reddit dataset relevant to the ongoing Russo-Ukrainian Conflict in 2022. Our dataset can support studies covering research areas such as social opinion analysis, politics characterization, and modeling of user relationships. We will keep collecting data from Reddit and updating this dataset.\footnote{\url{https://github.com/James-ZYM/RussiaUkraineConflict_Dataset}}

\section{Related Work}

\textbf{Online Social Networks datasets.} 
Researchers across various areas utilize Online Social Networks (OSNs) to study social and computing-related research questions such as polarisation and sentiment analysis. Several researchers share such datasets. Some of them aim to purely collect complete live data on social media, like Pushshift \cite{baumgartner2020pushshift}: a large-scale Reddit dataset including more than 600 million posts and 5 billion comments since June 2005; Snap \cite{snap}: a Twitter dataset containing 467 million Twitter posts from 20 million users covering seven months from 1 June 2009. 
In contrast, other OSNs datasets focus on specific topics or events, e.g., presidential elections \cite{elect1,elect2,allcott2017social}, Syrian War \cite{war1,war2}, and COVID-19 \cite{banda2021large,zarei2020first}. As for the Russo-Ukrainian Conflict, Cremisini released a dataset containing 4,538 news articles in English and performed bias detection for Pro-Ukraine and Pro-Russia stances \cite{cremisini2019challenging}. 

Most closely related are a small set of social media datasets pertaining to the Russo-Ukrainian Conflict, including Twitter \cite{chen2022tweets, haq2022twitter, pohl2022twitter}, VKontakte \cite{park2022voynaslov}, and Weibo \cite{fung2022weibo}. Reddit differs from these platforms due to its unique niche-oriented communities. Such a structure allows us to organize data in a more topic-centric way, supporting studies like language analysis \cite{reddit1}, user modeling \cite{haq2022short}, and sentiment analysis \cite{reddit2}. To our knowledge, there are no other Reddit datasets covering the Russo-Ukrainian Conflict. We present our dataset here to address this gap.

\textbf{Data analysis on crisis events.} Various researchers have studied crisis communication on OSNs in recent years. Saroj and Pal presented a systematic survey on the effect of emergencies on social media \cite{saroj2020use}. Their work discusses how post data can help crisis management. Nazan performed a comparative sentiment analysis on tweets related to the Syrian refugee crisis, revealing correlations between tweets' sentiment and the language of content \cite{ozturk2018sentiment}. 
For the COVID-19 crisis, Cinelli provided a comparative analysis of users’ activity on five different social media platforms and built a model to analyze information spread \cite{cinelli2020covid}. Regarding the ongoing Russo-Ukrainian Conflict, Hanley has analyzed Russian disinformation narratives utilizing semantic search \cite{hanley2022happenstance}. Such works depend on reliable and comprehensive datasets. In this work, we also perform a data characterization. We believe our results can serve as a reference for future studies in this area.

\section{Dataset}

\subsection{Reddit Data}

On Reddit, users can submit a post to start a new conversation on a specific topic by sharing text, links, or other multimedia. Accordingly, other users can comment or vote up/down the post. On Reddit, users subscribe to topic-specific communities called subreddits. Therefore, to collect data on a specific topic, several previous Reddit studies target particular subreddits \cite{weninger2013exploration,singer2014evolution}.

\subsection{Subreddit Selection}
\renewcommand{\arraystretch}{1.4}
\begin{table}[]
\centering
\resizebox{0.7\columnwidth}{!}{%
\begin{tabular}{|l|c|}
\hline
\textbf{Classes}  & \textbf{Key Words}                      \\ \hline
\textbf{Country} & \textit{\begin{tabular}[c]{@{}c@{}}Ukraine, Ukrainian, Russia, \\ Russian, Russo-Ukrainian\end{tabular}} \\ \hline
\textbf{Activity} & \textit{war, conflict, combat, defense} \\ \hline
\textbf{Leader} & \textit{Putin, Zelenskyy}               \\ \hline
\end{tabular}%
}
\caption{List of key words for searching subreddits.}
\label{tab:keywords}
\end{table}
\renewcommand{\arraystretch}{1}

We select relevant subreddits to collect posts and comments related to the conflict. As detailed in Table~\ref{tab:keywords}, we first create a list of keywords related to the Russo-Ukrainian Conflict.
We only consider English keywords because Reddit limits the names of subreddits to Latin characters, digits, and underscore.\footnote{\url{https://github.com/reddit-archive/reddit/blob/master/r2/r2/lib/validator/validator.py\#L512}}
In practice, we input each keyword into the search bar and record all the results in the \emph{Communities} section with at least 10,000 subscribers. In addition, we manually filter non-related subreddits based on their names and descriptions. We performed the search on 29 May 2022, discovering 67 subreddits potentially related to the Russo-Ukrainian Conflict. 
%We note that we did not perform further exploration and update the results. 
Note that any subreddits formed after 29 May 2022 are not included. Eventually, a subreddit is selected as a candidate once it belongs to any one of the following categories: 

\begin{enumerate}
    \item \textbf{R-U Conflict:} Subreddits directly focusing on the Russo-Ukrainian Conflict. These subreddits' names or community descriptions must include elements of the Russo-Ukrainian Conflict, e.g., Ukraine, Russia, and aliases for the Russo-Ukrainian Conflict. Posts and comments submitted on these subreddits focus on the crisis.
    
    \item \textbf{Military Related topic:} Subreddits containing general discussions on defense and military-related content, while recent posts were relevant to the Russo-Ukrainian Conflict. We manually check the first 50 posts in \emph{Hot}, \emph{Top}, and \emph{Rising} order respectively. If there are more than 30\% of posts collectively related to the conflict, this subreddit is viewed as relevant to the Russo-Ukrainian Conflict and selected.
    % \noteeh{how do you do this particular step?}
\end{enumerate}

Finally, we select 12 subreddits, as listed in Table~\ref{tab:seleted_subreddits}. We have also tried adding other potential topics that could generate new categories, e.g., politics and news. However, we find that the two categories above cover all the major potential subreddits. Other topics do not facilitate us getting any new related results.

\renewcommand{\arraystretch}{1.4}
\begin{table}[]
\resizebox{\columnwidth}{!}{%
\begin{tabular}{|l|l|}
\hline
\textbf{Category} & \textbf{Subreddit Name}                           \\ \hline
\multirow{4}{*}{\textbf{R-U Conflict}}     & \textit{ukraine, ukraina, UkraineConflict}   \\
                  & \textit{RussiaUkraineWar2022, UkrainianConflict}  \\
                  & \textit{UkraineWarReports, UkraineInvasionVideos} \\
                  & \textit{UkraineWarVideoReport}                    \\ \hline
\multirow{2}{*}{\textbf{Military Related}} & \textit{war, CombatFootage, CredibleDefense} \\
                  & \textit{geopolitics}                              \\ \hline
\end{tabular}%
}
\caption{Selected subreddits for each category (Results on 29 May 2022).}
\label{tab:seleted_subreddits}
\end{table}
\renewcommand{\arraystretch}{1}

\subsection{Data Collection}
\ehsan{We utilize pushshift.io API\footnote{\url{https://github.com/pushshift/api}} for data collection.}  We performed collection on 30 May 2022 and downloaded data from our 12 selected subreddits starting from 24 February 2022 to 29 May 2022. Additionally, to protect data integrity, we keep the data format consistent with the Pushshift dataset's \cite{baumgartner2020pushshift}, and preserve all columns in our final publicized dataset.

\begin{figure}[tp]
  \centering
  \includegraphics[width=\columnwidth]{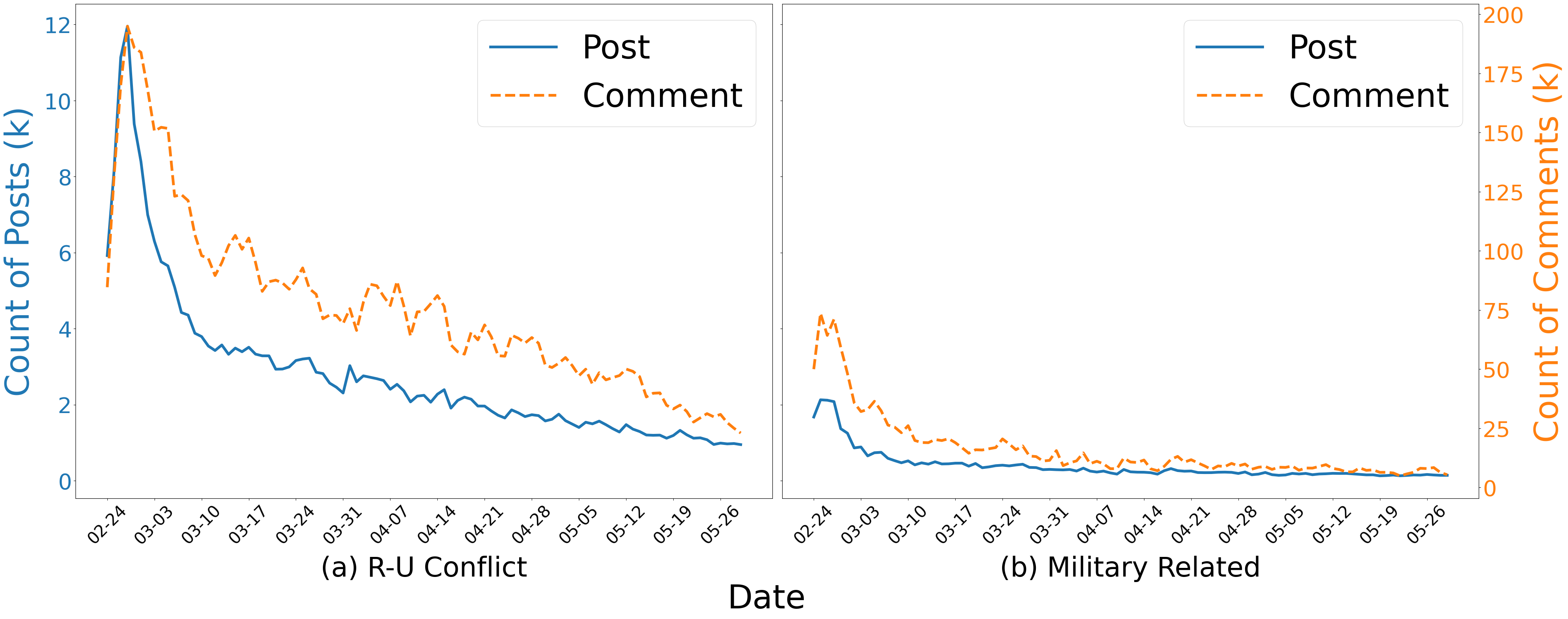}
  \caption{Daily volume of posts and comments for each category; (a): R-U Conflict; (b): Military Related.}
  \label{fig:cnt_cat}
\end{figure}

As a result, by 30 May,\footnote{30 May 2022 0:00:00 UTC} we have collected 306,482 posts and 8,751,705  comments from the selected subreddits. Figure~\ref{fig:cnt_cat} shows the distribution of daily data volume for each category. Around 95,349 items are contributed to all these subreddits each day. 

\section{Data Analysis}

\subsection{Descriptive Analysis}

This section depicts the metadata in each category. 
% The analysis results reflect a general situation of subscribers' activity in different subreddits. 
Table~\ref{tab:overall_cnt} shows overall frequencies.

\renewcommand{\arraystretch}{1.3}
\begin{table*}[]
\resizebox{\textwidth}{!}{%
\begin{tabular}{|llccccccc|}
\hline
\multicolumn{1}{|l|}{\textbf{Category}} &
  \textbf{Subreddit Name} &
  \textbf{Created Date} &
  \textbf{\#Subscribers(k)} &
  \textbf{\#Active Users} &
  \textbf{\#Total Posts} &
  \textbf{\#Total Comments} &
  \textbf{Avg \#Replies/Post} &
  \textbf{Avg Scores/Post} \\ \hline
\multicolumn{1}{|l|}{\multirow{9}{*}{\textbf{R-U Conflict}}} &
  \textit{ukraine} &
  Dec 23, 2008 &
  731 &
  269,612 &
  125,008 &
  3,800,366 &
  0.72 &
  1.58 \\
\multicolumn{1}{|l|}{} &
  \textit{ukraina} &
  Mar 14, 2014 &
  119 &
  23,481 &
  11,058 &
  100,572 &
  0.36 &
  1.12 \\
\multicolumn{1}{|l|}{} &
  \textit{RussiaUkraineWar2022} &
  Feb 24, 2022 &
  105 &
  37,546 &
  15,788 &
  275,837 &
  0.54 &
  1.14 \\
\multicolumn{1}{|l|}{} &
  \textit{UkrainianConflict} &
  Feb 20, 2014 &
  354 &
  122,278 &
  56,086 &
  1,437,688 &
  0.49 &
  1.45 \\
\multicolumn{1}{|l|}{} &
  \textit{UkraineConflict} &
  Mar 2, 2014 &
  26 &
  10,801 &
  8,899 &
  54,020 &
  0.06 &
  1.05 \\
\multicolumn{1}{|l|}{} &
  \textit{UkraineWarVideoReport} &
  Feb 24, 2022 &
  428 &
  150691 &
  34,989 &
  1,394,659 &
  0.56 &
  1.36 \\
\multicolumn{1}{|l|}{} &
  \textit{UkraineWarReports} &
  Feb 24, 2022 &
  55.6 &
  16,134 &
  7,242 &
  77,787 &
  0.61 &
  1.09 \\
\multicolumn{1}{|l|}{} &
  \textit{UkraineInvasionVideos} &
  Feb 24, 2022 &
  51.1 &
  16,119 &
  10,110 &
  91,716 &
  0.02 &
  1.06 \\
\multicolumn{1}{|l|}{} &
  \textbf{Total} &
  \textbf{-} &
  \textbf{1750.7} &
  \textbf{467,352} &
  \textbf{269,180} &
  \textbf{7,232,645} &
  \textbf{0.57} &
  \textbf{1.43} \\ \hline
\multicolumn{1}{|l|}{\multirow{5}{*}{\textbf{Military Related}}} &
  \textit{war} &
  Mar 19, 2008 &
  92.5 &
  18,018 &
  9,167 &
  97,119 &
  0.45 &
  1.03 \\
\multicolumn{1}{|l|}{} &
  \textit{CombatFootage} &
  Sep 10, 2012 &
  1,035 &
  127,638 &
  24,765 &
  1,287,740 &
  1.10 &
  1.26 \\
\multicolumn{1}{|l|}{} &
  \textit{CredibleDefense} &
  Aug 9, 2013 &
  58.6 &
  6,780 &
  1,090 &
  91,536 &
  0.65 &
  1.03 \\
\multicolumn{1}{|l|}{} &
  \textit{geopolitics} &
  Mar 26, 2008 &
  435 &
  8,407 &
  2,280 &
  42,665 &
  1.48 &
  1.03 \\
\multicolumn{1}{|l|}{} &
  \textbf{Total} &
  \textbf{-} &
  \textbf{1621.1} &
  \textbf{150,489} &
  \textbf{37,302} &
  \textbf{1,519,060} &
  \textbf{0.95} &
  \textbf{1.18} \\ \hline
\multicolumn{2}{|c|}{\textbf{All}} &
  \textbf{-} &
  \textbf{2936.8} &
  \textbf{546,593} &
  \textbf{306,482} &
  \textbf{8,751,705} &
  \textbf{0.62} &
  \textbf{1.40} \\ \hline
\end{tabular}%
}
\caption{Frequencies of metadata for the selected subreddits (Results are based on data collected from 24 Feb 2022 to 29 May 2022. We performed the data collection on 30 May 2022). The \enquote{Active Users} refer to those users who ever sent at least one post(s) or comment(s) during the data collection period.}

\label{tab:overall_cnt}
\end{table*}
\renewcommand{\arraystretch}{1}

\subsubsection{Daily volume of posts and comments}

To understand the trend of subscribers' activity during this crisis event, Figure~\ref{fig:cnt_cat}. presents the time series of each category's submitted posts and comments. 
We observe that posts and comments in \enquote{R-U Conflict} constitute the vast majority of the total volume. As for the distribution, the two categories' volumes of posts and comments spiked to more than 10,000 and 230,000 per day during the first week after the crisis. After that, the volumes decreased dramatically during the second week and gradually flattened out in the end. This suggests that the Russo-Ukrainian Conflict spread rapidly in these communities after the war broke out. The attention of those subscribers then shifted away quickly after two weeks. 

\subsubsection{Engagement of posts}

\begin{figure}[tp]
  \centering
  \includegraphics[width=\columnwidth]{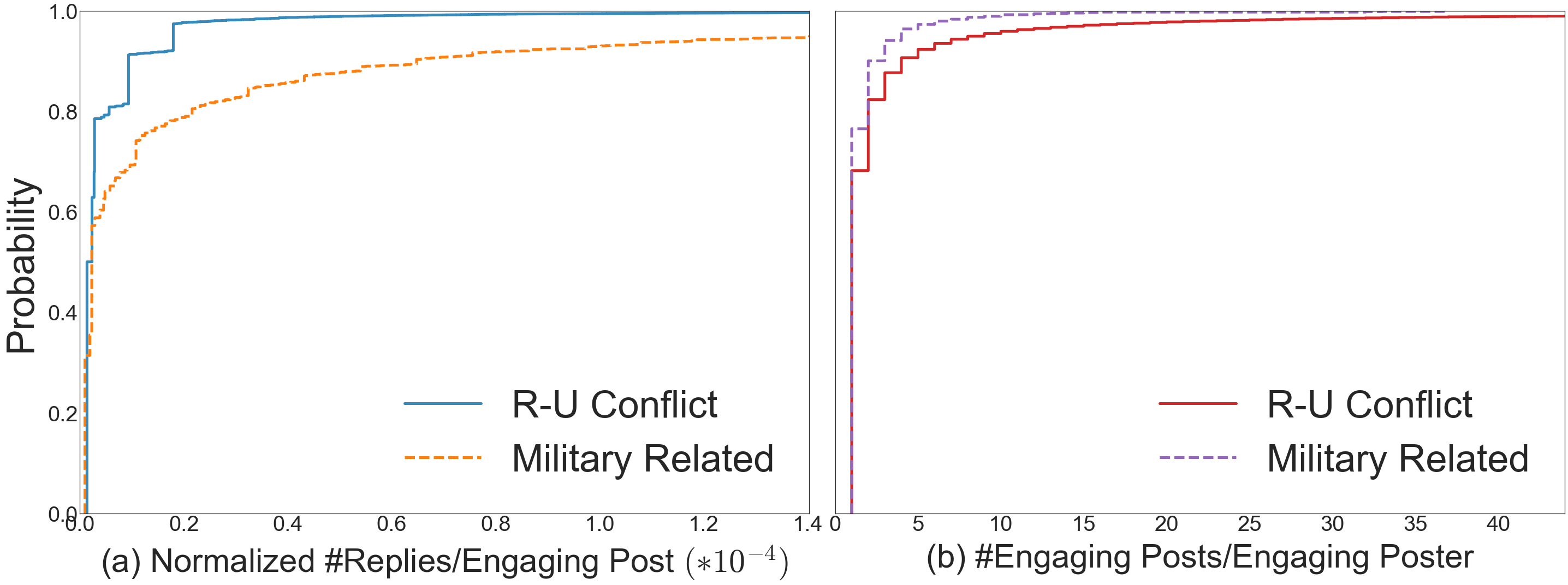}
  \caption{(a): CDF plot for normalized replies number per \emph{engaging post}, replies number divided by the size of subscribers for each subreddit; (b): CDF plot for number of \emph{engaging posts} per \emph{engaging} poster (who has submitted at least one \emph{engaging posts} before).}
  \label{fig:cdf_replies}
\end{figure}

We further examine the engagement of both categories. Prior studies have revealed a high correlation between the topic continuity and engagement of posts \cite{zeng2021modeling,pethe2019trumpiest}. Especially when evaluating the engagement between users, the number of replies (comments generated under a post) works as a significant predictive indicator.
More replies mean more engagement among users, indicating the topic will continue in further discussion \cite{haq2022short, wang2022successful}.

In addition, we conjecture that the number of subscribers in a subreddit might contribute to the likelihood of a post receiving replies. Here, we use the normalized number of replies to evaluate the engagement between the poster and commenters (i.e., we divide the number of replies by the number of subscribers in a subreddit). We first notice a large number of posts with no replies, introducing a significant bias when examining the distribution of replies. Overall, only 27\% of posts have received replies during the collection period, implying that a small amount of posts generates the most engagement in each category. Therefore, to concentrate our analysis on those posts driving users' engagement, the calculation and descriptive statistics are only based on posts with at least one reply. We refer such posts as \emph{engaging posts}.
Figure~\ref{fig:cdf_replies}(a) shows the distribution of their normalized number of replies. Although \enquote{Military Related} subreddits have fewer posts in total, according to the statistics of normalized replies number, their engaging posts receive more replies on average ($min=0.010, max=61.749, \mu=0.361, median=0.023$, and $\sigma = 1.865$) than those in \enquote{R-U Conflict} ($min=0.014, max=29.904, \mu=0.057, median=0.014$, and $\sigma = 0.295$), reflecting more substantial engagement between posters and commenters. In addition, we also find that replies of engaging posts for \enquote{R-U Conflict} follow a heavy-tailed distribution.

Previous study shows that discussions in a subreddit with high engagement are mainly generated by certain groups of users \cite{choi2015characterizing}. \james{%Inspired by their results, 
Thus, we align with the prior approach and also refer the submitters of engaging posts as \emph{engaging posters}. Furthermore, we analyze the distribution of the number of engaging posts they have submitted, as shown in Figure~\ref{fig:cdf_replies}(b). 
% Compared to \enquote{Military Related} ($min=1, max=37, \mu=1.58, median=1.00$, and $\sigma = 2.03$), engaging posters in \enquote{R-U Conflict} have submitted more engaging posts on average ($min=1, max=1857, \mu=3.45, median=1.00$, and $\sigma = 20.09$) during the collection period.
} 
% \gareth{Could we add a brief description of this figure. What does the X-axis mean by 'interactive posts'? Does this mean engaging posts?}
Furthermore, only \james{12\%}
% \gareth{Minor point - would be nice to avoid the word 'around'. It takes up space and conveys an impression of fuzziness. Instead, we could just say 10\% (or 10.x\% if we want to be more precise.} 
of users in this category have ever generated engaging posts, significantly lower than \enquote{R-U Conflict} (36\%). 

\subsubsection{Rank of posts}

\begin{figure}[]
  \centering
  \includegraphics[width=\columnwidth]{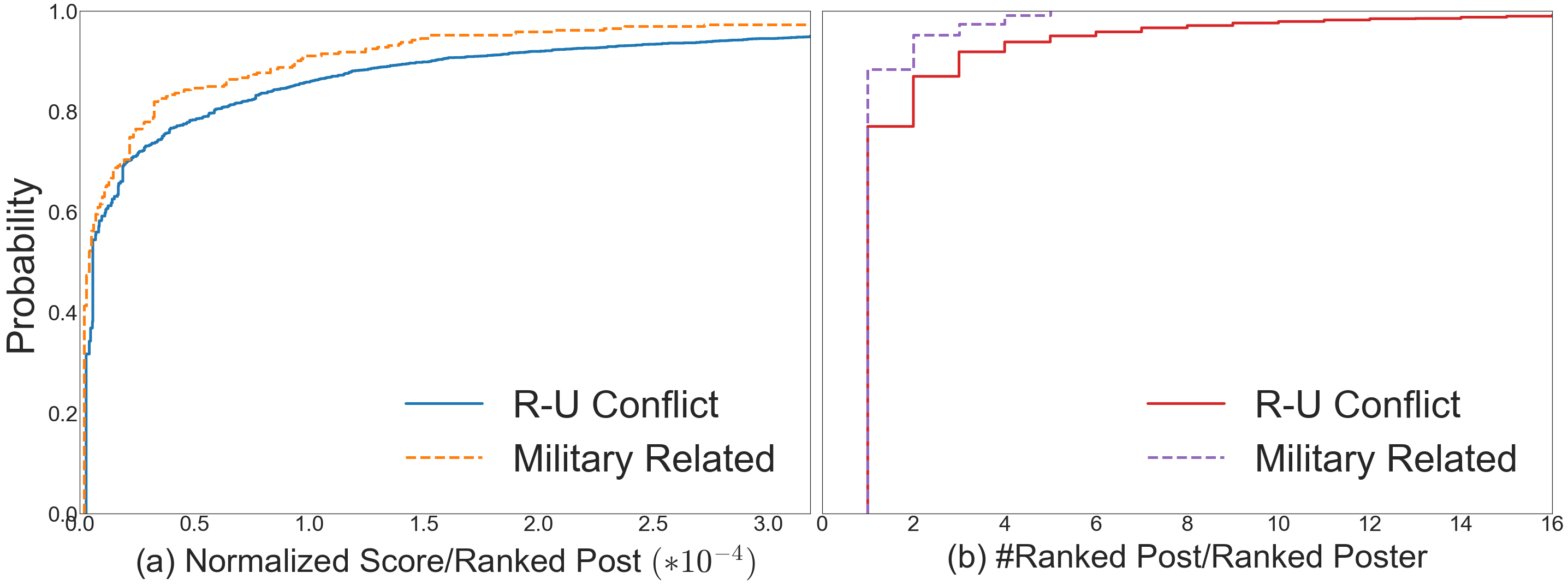}
  \caption{(a): CDF plot for normalized scores in \emph{ranked posts}, score divided by the size of subscribers for each subreddit; (b): CDF plot for number \emph{ranked posts} per \emph{ranked poster} (who has submitted at least one \emph{ranked posts} before).}
  \label{fig:cdf_score}
\end{figure}

We next explore posts with a high rank in \james{the observed subreddits, an aggregated evaluation of posts' priority on the \emph{Hot} section. Posts' ranks are influenced by vote scores.}
% \gareth{What do we mean by category here? The two groups of subreddits?}
After a post is submitted, viewers can up-vote or down-vote it. Generally, posts with higher vote scores (up-votes - down-votes) are ranked higher than posts with lower vote scores. Several previous studies have utilized vote scores in works like user behavior studies \cite{weninger2015random} and discussion threads analysis \cite{weninger2013exploration}.

Here we investigate the distribution of highly ranked posts' scores. Since Reddit automatically makes submitters up-vote their posts, the default score of all newly submitted posts is 1. Similarly, we also notice a large number of posts with default scores, introducing a strong bias in distribution analysis. \james{Similar to the previous section, to concentrate on posts with a higher rank, we refer those posts with score $>$ 1 as \emph{ranked posts} and their submitters as \emph{ranked posters}. We also draw the distribution plots for normalized score of ranked posts (i.e., we divide the vote score by the number of subscribers in a subreddit) and the number of such posts for their ranked posts, see Figure~\ref{fig:cdf_score}. According to the statistics of normalized score, 
% both two categories present similar distribution of scores received by ranked posts. 
compared to \enquote{Military Related} ($min=0.019, max=15.652, \mu=0.372, median=0.039$, and $\sigma = 1.159$), ranked posts are more likely to receive higher vote scores in \enquote{R-U Conflict} ($min=0.027, max=58.22, \mu=0.742, median=0.056$, and $\sigma = 2.716$). 
% Furthermore, ranked posters in \enquote{R-U Conflict} ($min=1, max=82, \mu=1.91, median=1.00$, and $\sigma = 3.93$) have also submitted more ranked posts than those in \enquote{Military Related} ($min=1, max=5, \mu=1.20, median=1.00$, and $\sigma = 0.64$).
} 
% \gareth{Anything explicit to flag here?}

\subsection{Text Content}

We next explore the characteristics of users' language usage and opinions. Table~\ref{tab:overall_txt} shows the average of linguistic features in our dataset.

\renewcommand{\arraystretch}{1.1}
\begin{table*}[]
\resizebox{\textwidth}{!}{%
\begin{tabular}{|llccccc|}
\hline
\multicolumn{1}{|l|}{\textbf{Category}} &
  \textbf{Subreddit Name} &
  \textbf{Title Length/Post} &
  \textbf{Text Length/Post} &
  \textbf{Text Length/Comment} &
  \textbf{\#Emojis/Post} &
  \textbf{\#Emojis/Comment} \\ \hline
\multicolumn{1}{|l|}{\multirow{9}{*}{\textbf{R-U Conflict}}} &
  \textit{ukraine} &
  13.39 &
  21.76 &
  29.17 &
  0.115 &
  0.060 \\
\multicolumn{1}{|l|}{} &
  \textit{ukrain} &
  13.24 &
  18.05 &
  24.29 &
  0.171 &
  0.152 \\
\multicolumn{1}{|l|}{} &
  \textit{RussiaUkraineWar2022} &
  14.31 &
  9.94 &
  28.98 &
  0.177 &
  0.106 \\
\multicolumn{1}{|l|}{} &
  \textit{UkrainianConflict} &
  15.91 &
  0.32 &
  30.40 &
  0.055 &
  0.043 \\
\multicolumn{1}{|l|}{} &
  \textit{UkraineConflict} &
  15.05 &
  3.95 &
  30.22 &
  0.102 &
  0.085 \\
\multicolumn{1}{|l|}{} &
  \textit{UkraineWarVideoReport} &
  12.96 &
  10.72 &
  25.27 &
  0.061 &
  0.064 \\
\multicolumn{1}{|l|}{} &
  \textit{UkraineWarReports} &
  13.08 &
  10.52 &
  27.79 &
  0.091 &
  0.077 \\
\multicolumn{1}{|l|}{} &
  \textit{UkraineInvasionVideos} &
  13.51 &
  4.47 &
  21.97 &
  0.093 &
  0.128 \\
\multicolumn{1}{|l|}{} &
  \textbf{Total} &
  \textbf{13.96} &
  \textbf{12.96} &
  \textbf{28.45} &
  \textbf{0.098} &
  \textbf{0.061} \\ \hline
\multicolumn{1}{|l|}{\multirow{5}{*}{\textbf{Military Related}}} &
  \textit{war} &
  11.71 &
  0.67 &
  26.23 &
  0.065 &
  0.063 \\
\multicolumn{1}{|l|}{} &
  \textit{CombatFootage} &
  11.87 &
  3.41 &
  25.72 &
  0.025 &
  0.024 \\
\multicolumn{1}{|l|}{} &
  \textit{CredibleDefense} &
  16.15 &
  9.70 &
  62.42 &
  0.009 &
  0.006 \\
\multicolumn{1}{|l|}{} &
  \textit{geopolitics} &
  12.15 &
  3.63 &
  62.56 &
  0.022 &
  0.008 \\
\multicolumn{1}{|l|}{} &
  \textbf{Total} &
  \textbf{11.97} &
  \textbf{2.92} &
  \textbf{29.08} &
  \textbf{0.034} &
  \textbf{0.025} \\ \hline
\multicolumn{2}{|c|}{\textbf{All}} &
  \textbf{13.72} &
  \textbf{11.80} &
  \textbf{28.56} &
  \textbf{0.091} &
  \textbf{0.055} \\ \hline
\end{tabular}%
}
\caption{Average of linguistic features for the selected subreddits, (based on data collected from 24 Feb 2022 to 29 May 2022).}
\label{tab:overall_txt}
\end{table*}
\renewcommand{\arraystretch}{1}

\subsubsection{Term frequency}
% word cloud
Several related works on Reddit show that term frequency is an essential factor in content analysis, e.g., semantic content analysis related to weight loss \cite{pappa2017factors} and politics characterization for Reddit communities \cite{soliman2019characterization}. Term frequency also reflects the distribution of language usage, revealing the trend of users' topics, opinions, or attitudes.

\begin{figure*}[]
     \centering
     \begin{subfigure}[b]{0.4\textwidth}
         \centering
         \includegraphics[width=\textwidth]{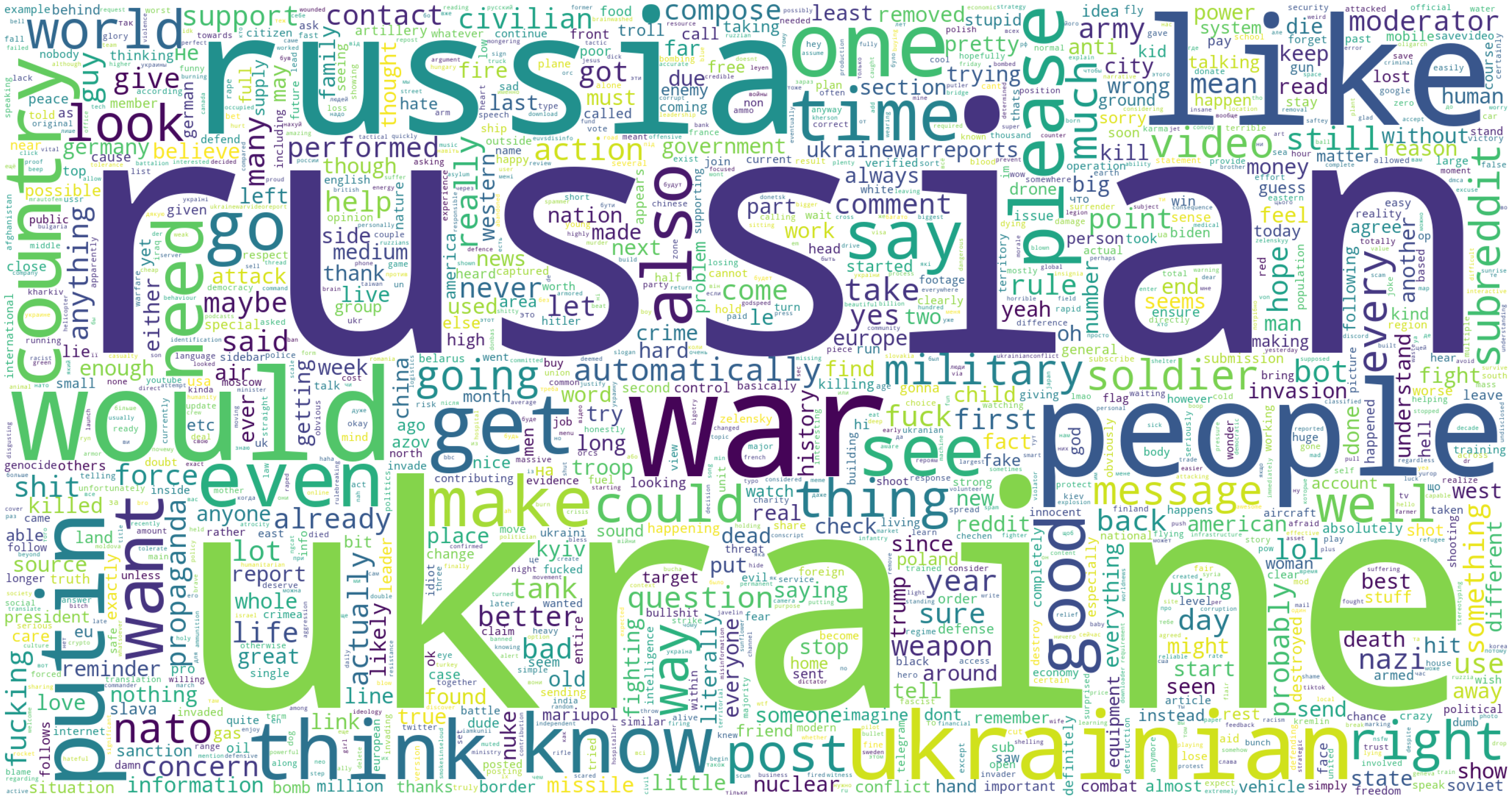}
         \caption{R-U Conflict (most used terms)}
         \label{fig:wc_ru}
     \end{subfigure}
    %  \hfill
     \begin{subfigure}[b]{0.4\textwidth}
         \centering
         \includegraphics[width=\textwidth]{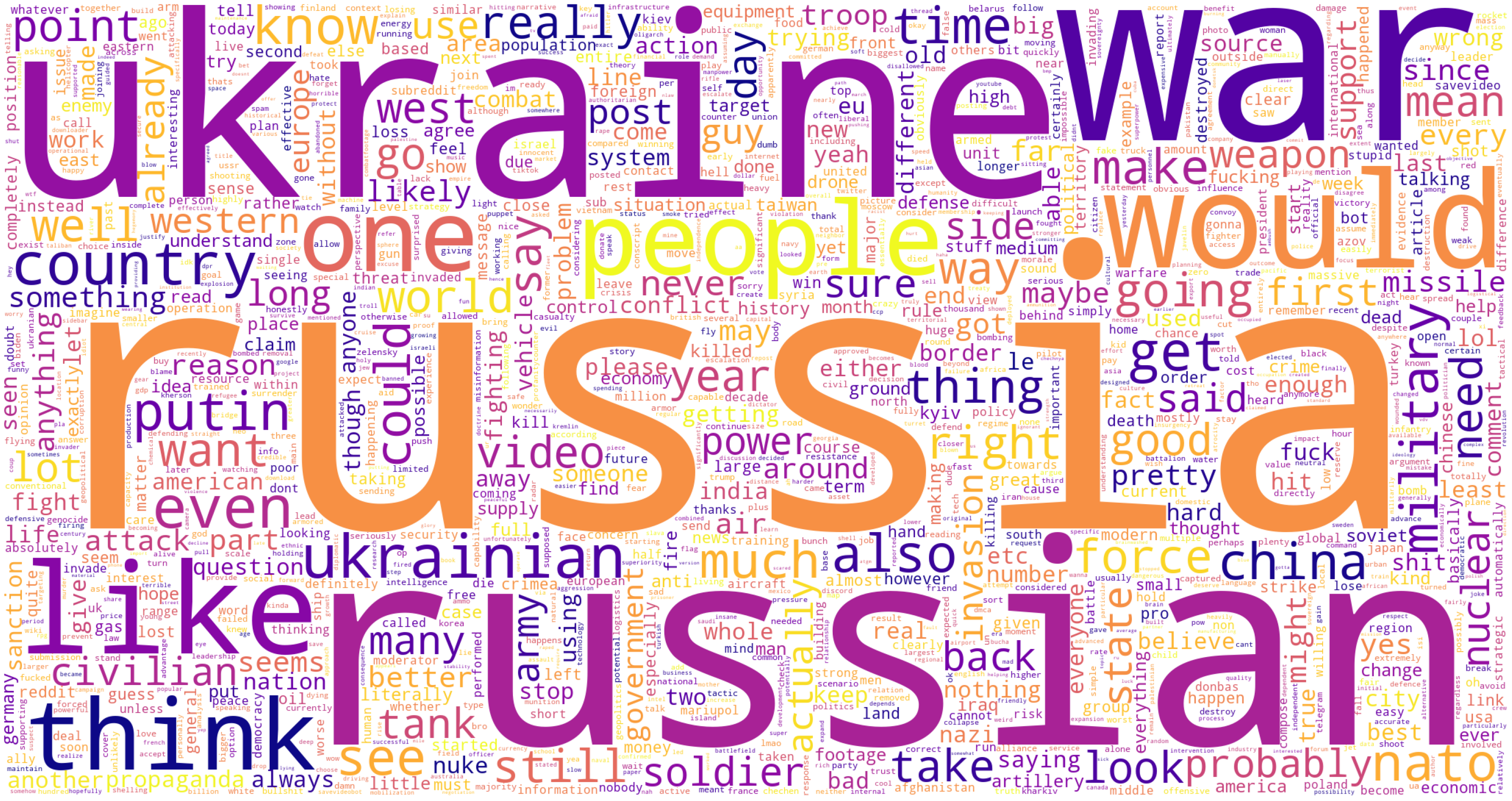}
         \caption{Military Related (most used terms)}
         \label{fig:wc_mr}
     \end{subfigure}
    %  \hfill
     \begin{subfigure}[b]{0.4\textwidth}
         \centering
         \includegraphics[width=\textwidth]{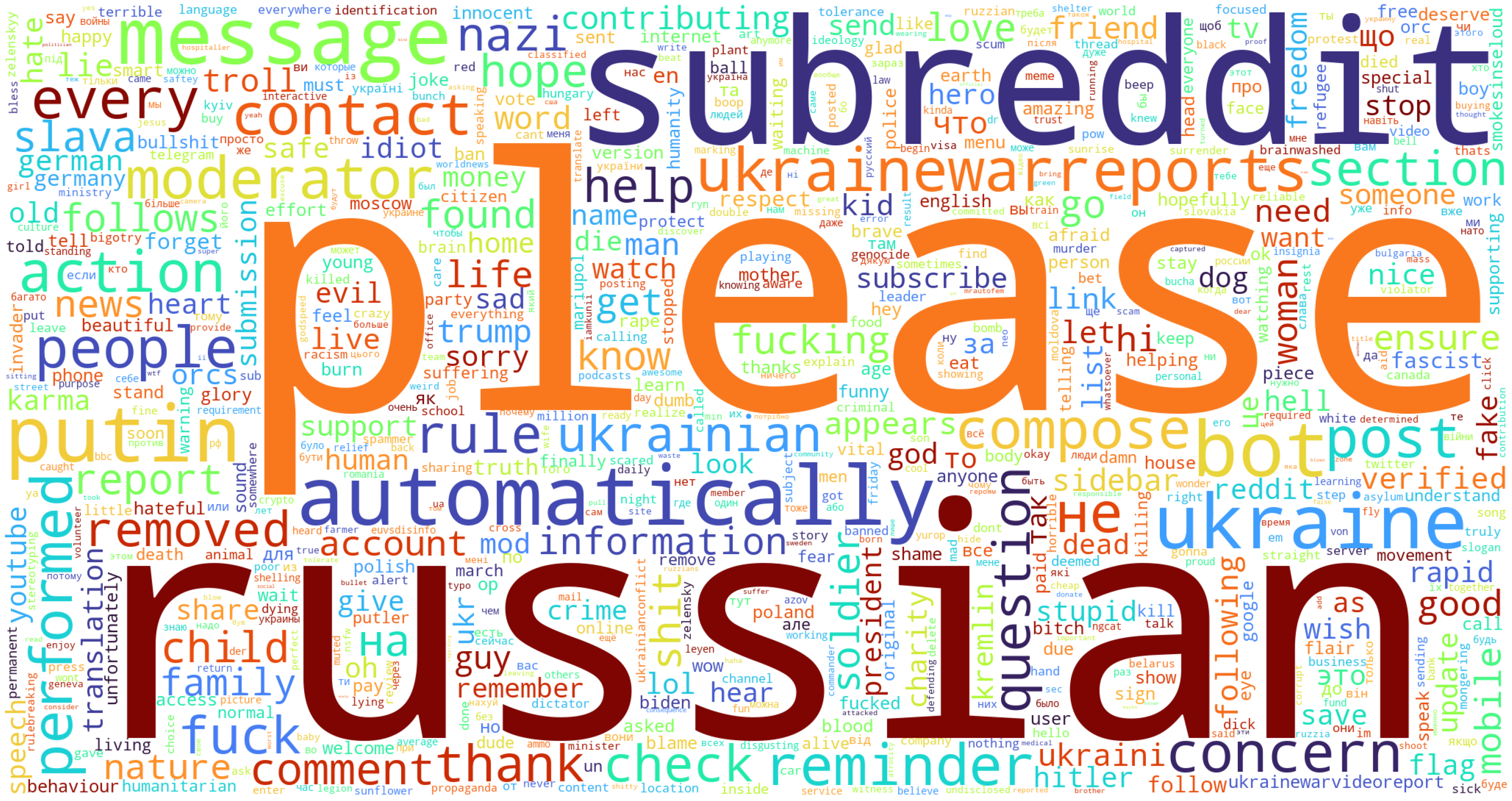}
         \caption{R-U Conflict (characteristic terms)}
         \label{fig:wc_ru_key}
     \end{subfigure}
    %  \hfill
     \begin{subfigure}[b]{0.4\textwidth}
         \centering
         \includegraphics[width=\textwidth]{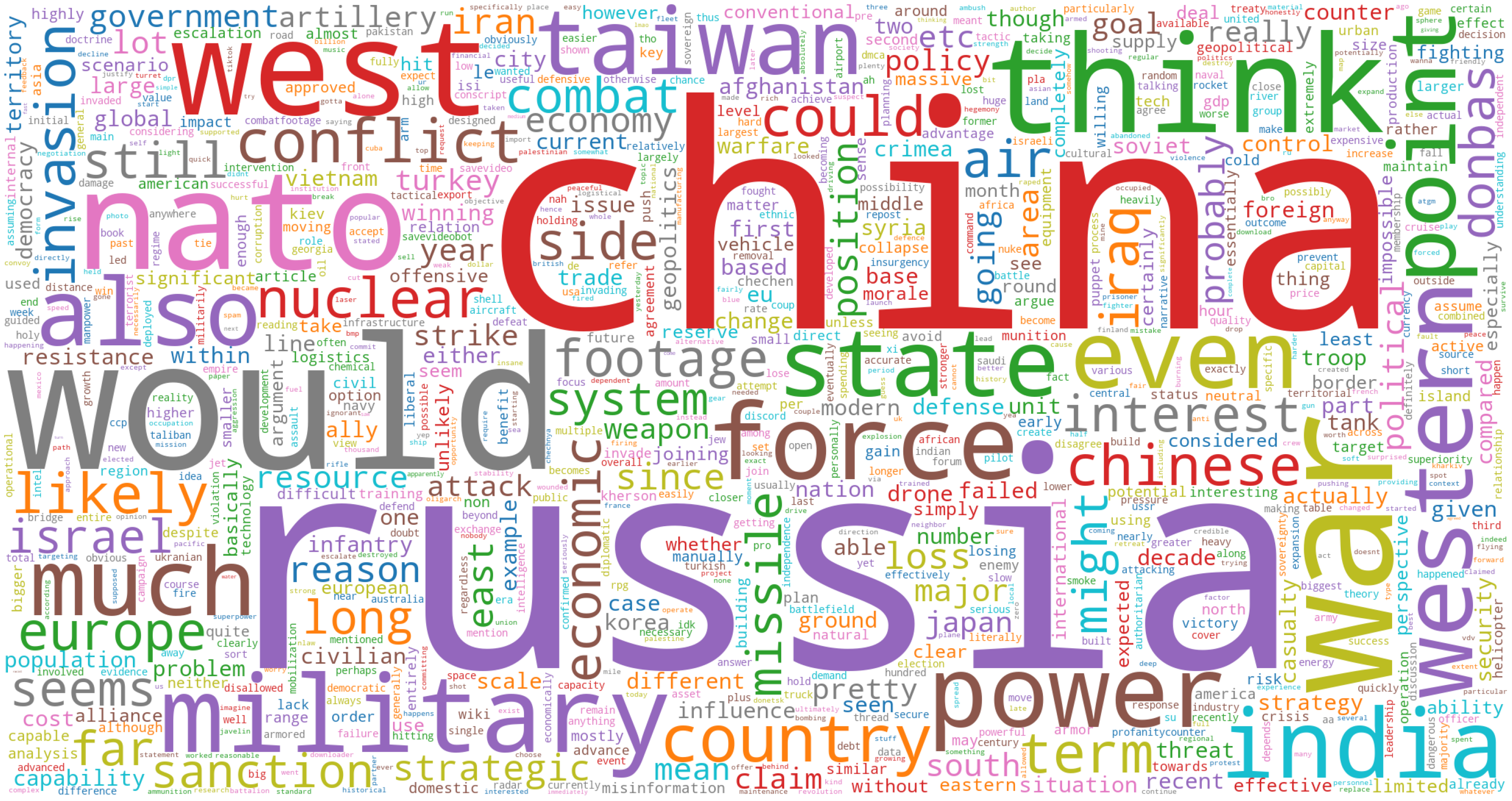}
         \caption{Military Related (characteristic terms)}
         \label{fig:wc_mr_key}
     \end{subfigure}
        \caption{Most used and characteristic terms of categories; In (a) and (b) we plot the union sets for top 1,000 most used terms of each subreddis in \enquote{R-U Conflict} and \enquote{Military Related} respectively; In (c) and (d) we plot the differences of normalized term frequency for words in each category's union set compared to the other.
opposed subreddit}
        \label{fig:wc_all}
\end{figure*}

For each subreddit, examine the term frequency (i.e., the count of a term divided by the total number of terms in the subreddit). Then, we select the top 1K terms with the highest frequency in each subreddit and create a union set for them. We define the terms for each union set as the most used terms in each category. After that, we calculate the normalized term frequency (i.e., for a term in one union set, the average value of all its term frequencies for subreddits in this category). Figure~\cref{fig:wc_ru,fig:wc_mr} show the word clouds of these terms weighted by the normalized term frequency after lemmatization for nouns and stop word removal.

We further calculate the relative difference of the normalized term frequency \cite{soliman2019characterization} in the two categories. For \enquote{R-U Conflict}, we subtract the normalized term frequency of relative words in \enquote{Military Related}, and vice versa. 
This allows us to discern the terms used more frequently in one category than the other. \james{For each category, we refer such terms as its \emph{characteristic terms}.} We utilize relative differences to weight characteristic terms in word clouds, see \cref{fig:wc_ru_key,fig:wc_mr_key}.
% \gareth{How do we formally define characteristic terms? I think the previous sentence might need to be rewritten.}
According to the plots, the two categories show different stances toward the ongoing war. In \enquote{R-U Conflict}, users tend to use terms with higher negative connotations, e.g., \emph{orcs}, \emph{evil}, \emph{idiot}, \emph{troll}, and F-words. 
We consider that such patterns for characteristic terms, compared to \enquote{Military Related}, reveal a preference for the Pro-Ukraine stance in \enquote{R-U Conflict}. In contrast, \enquote{Military Related} subscribers prefer terms highly related to global politics and the military. These terms contain much more normal connotations, indicating a neutral stance in this category compared to \enquote{R-U Conflict}.

\subsubsection{Linguistic features}

Stylometric properties, including text length and emojis use, have been widely applied in user profiling and accounts classification tasks \cite{lagutina2019survey,kosmajac2020twitter}. To understand the distribution of these properties in users’ discussions, we measure the average text length and number of emojis for the main text (both posts and comments). 

% According to Table~\ref{tab:overall_txt}, we observe that, instead of the categories, the statistics values are highly related to the properties of subreddits, e.g., topic choices and community rules for the post format. For example, \emph{UkraineWarVideoReport} and \emph{UkraineInvasionVideos} both belong to \enquote{R-U Conflict}, but the latter has a shorter text length on average for the popularity of posting videos in this community. 

\begin{figure*}[]
    \centering
    \includegraphics[width=.75\textwidth]{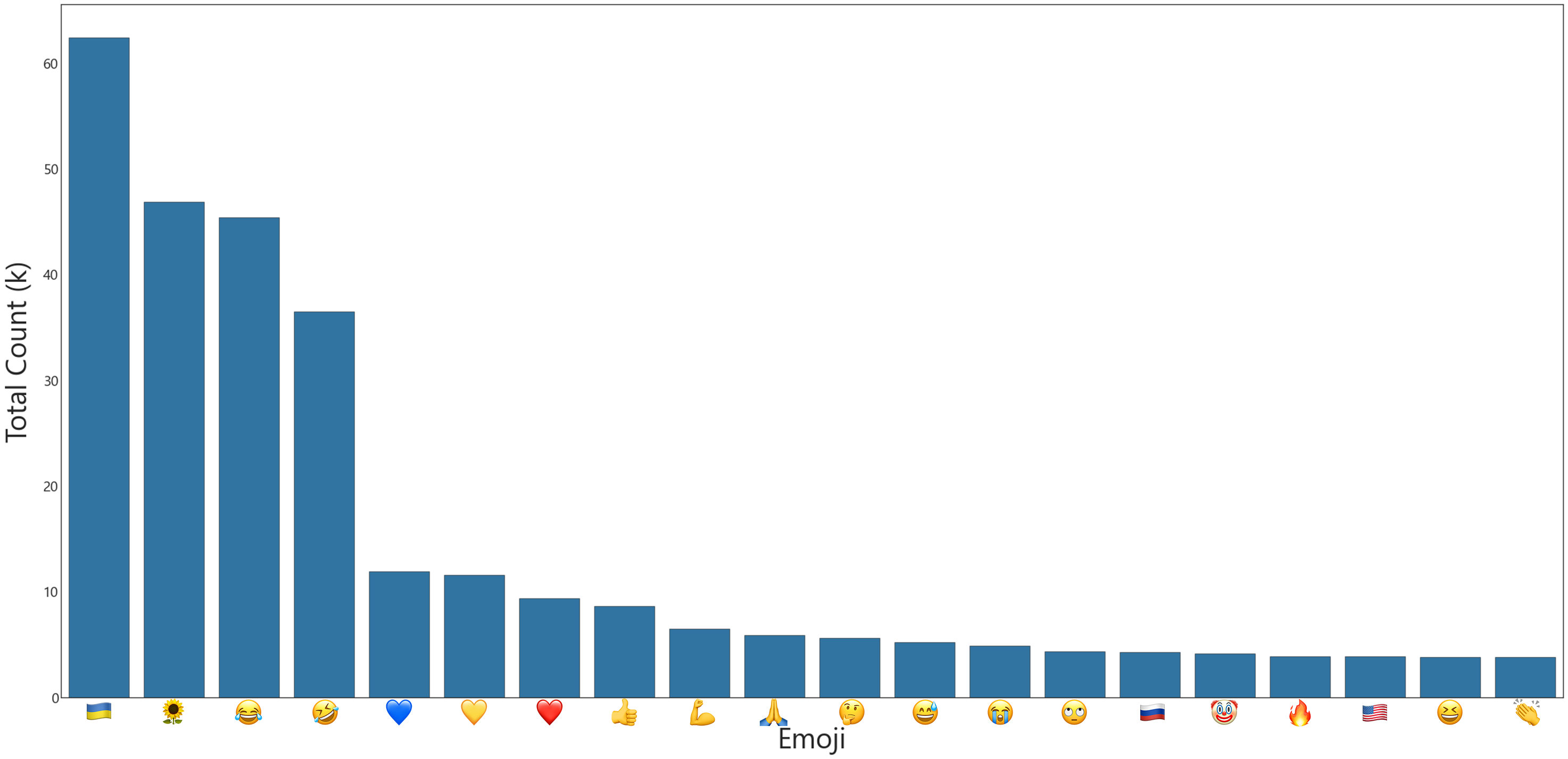}
    \caption{\james{Count} of the top 20 most frequently used emojis in the whole dataset. 
    % \gareth{I guess this caption is wrong unless you are adding another graph? Also, what is the meaning of the colour code in the graph? If the colours have no meaning, I would set all the bars to the same colour.}
    }
    \label{fig:emoji}
\end{figure*}

In the online discussion on social platforms, specific patterns of widely used emoji phrases can reflect the user behaviors and public attitudes to ongoing topics, e.g., \includegraphics[width=1.1em]{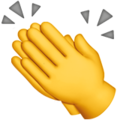}: Applause and agreement on the topic, or \includegraphics[width=1.1em]{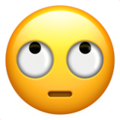}: Moderate disdain, disapproval, frustration. Aiming to gain an overview of the emojis usage patterns in the whole dataset, we present the usage count for the most popular emojis in Figure~\ref{fig:emoji}. We notice that there is a strong bias toward the Pro-Ukraine stance in emoji usage, implied by frequently used phrases like \emph{\enquote{Ukrainian heart}} (\includegraphics[width=1.1em]{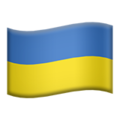}\includegraphics[width=1.1em]{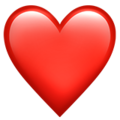} or \includegraphics[width=1.1em]{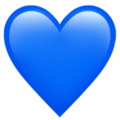}\includegraphics[width=1.1em]{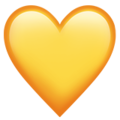}), \emph{\enquote{sunflower seed}} (\includegraphics[width=1.1em]{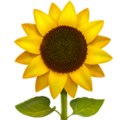})\footnote{\url{https://www.scmp.com/news/world/russia-central-asia/article/3168742/how-sunflowers-became-symbol-ukraines-resistance}} and \emph{\enquote{Russian clown}} (\includegraphics[width=1.1em]{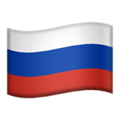}\includegraphics[width=1.1em]{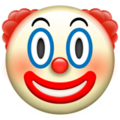}). For context, here is an overt example of a post using emoji phrases in the Pro-Ukraine stance: 

\begin{displayquote}
\emph{\#ukrainestrong! SLAVA UKRAINE! \includegraphics[width=1.1em]{figs/blue-heart.png}\includegraphics[width=1.1em]{figs/yellow-heart.png} if any aggressors see her!! well you now a sunflower!!! \includegraphics[width=1.1em]{figs/blue-heart.png}\includegraphics[width=1.1em]{figs/yellow-heart.png}\includegraphics[width=1.1em]{figs/sunflower.png}}
\end{displayquote}

% \begin{figure*}[]
%      \centering
%      \begin{subfigure}[b]{0.85\textwidth}
%          \centering
%          \includegraphics[width=\textwidth]{figs/lin_sta.png}
%          \caption{Average number of text length and emojis for subreddits.}
%          \label{fig:lin_sta}
%      \end{subfigure}
%      \hfill
%      \begin{subfigure}[b]{0.85\textwidth}
%          \centering
%          \includegraphics[width=\textwidth]{figs/emoji.png}
%          \caption{Top 20 popular emojis in the whole dataset.}
%          \label{fig:emoji}
%      \end{subfigure}
%         \caption{Statistics of text length and emojis; (a) measures the average count for length and emojis number of text contents, both posts and comments, in each selected subreddits; (b) shows the top 20 most frequently used emojis in the whole dataset. }
%         \label{fig:lin_all}
% \end{figure*}

% \subsubsection{Domain}
% leave this part first, since no very valuable results found.
% top popular domain (need to exclude Reddit itself?)

\subsubsection{Sentiment}

Finally, we investigate the changes in users' sentiment in posts. We utilize the VADER sentiment analysis tool \cite{hutto2014vader} to classify posts' text (including the title and the corresponding main text). We perform the sentiment classification based on VADER compound scores, an aggregated score reflecting the sentiment bias. A related sentiment analysis for Twitter text has come up with an effective standard \cite{elbagir2019twitter}, and we here apply the same strategy: \textbf{Positive:} VADER compound score $>$ 0.001; \textbf{Negative:} VADER compound score $<$ 0.001; \textbf{Neutral:} Otherwise.

We calculate the normalized frequency for each sentiment class in every category by date. We first examine the relative frequency for each sentiment class in every subreddit (the ratio of the number of posts in a class vs.\ the total number of posts in a subreddit). Then, the normalized frequency for a class in each category is the average value of all its relative frequencies. 

\begin{figure*}[]
    \centering
    \includegraphics[width=.7\textwidth]{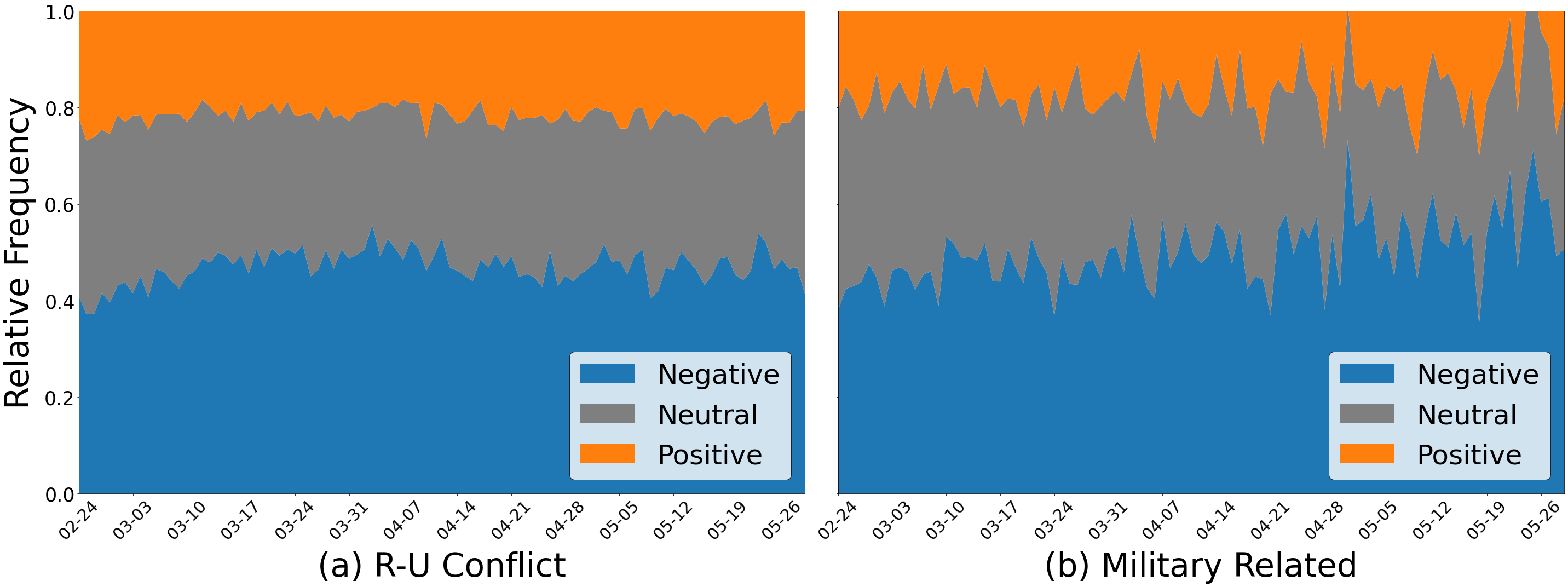}
    \caption{Distribution of daily users' sentiment in posts; (a) Sentiment distribution of posts in \enquote{R-U Conflict}; (b) Sentiment distribution of posts in \enquote{Military Related}.}
    \label{fig:sentiment}
\end{figure*}

Figure~\ref{fig:sentiment} shows the distribution of daily sentiment based on the normalized frequency. We observe that \enquote{R-U Conflict} ($min=0.37, max=0.56, \mu=0.47, median=0.47$, and $\sigma = 0.04$) and \enquote{Military Related} ($min=0.35, max=0.73, \mu=0.50, median=0.50$, and $\sigma = 0.07$) have a similar ratio of \emph{negative} class, but subreddits in \enquote{Military Related} ($min=0.09, max=0.30, \mu=0.18, median=0.18$, and $\sigma = 0.04$) contains relatively fewer \emph{positive} posts than in \enquote{R-U Conflict} ($min=0.18, max=0.27, \mu=0.22, median=0.22$, and $\sigma = 0.02$). In addition, \enquote{Military Related} subreddits carry larger variations on its distribution of \emph{negative} posts ratio, especially after April 21. 

% We try associating this trend with the ongoing war's timeline, and reveal a potential correlation between ratio changes and events, e.g., \enquote{canyon} before April 21 (Battle of Donbas\footnote{\url{https://en.wikipedia.org/wiki/Battle_of_Donbas_(2022)}}), surges after April 28 (2022 Transnistria attacks\footnote{\url{https://en.wikipedia.org/wiki/2022_Transnistria_attacks}}).

\section{Discussion \& Conclusion}

Social media datasets play an essential role in exploring research questions related to society. We shortlist 12 prominent subreddits relevant to the Russo-Ukrainian Conflict via an exhaustive subreddit exploration. Our dataset has collected more than 300K posts and 8M comments from these selected subreddits. 
To shed light on the dataset's features, we have also performed several characterization analyses on daily volumes, posts' engagement, rank, term frequency, linguistic features, and the sentiment of text. The ongoing Russo-Ukrainian Conflict has been discussed widely on Reddit since the first week of the war. After that, the daily volume decreased significantly and flattened out within one month, implying a shift away in Reddit users' attention. Furthermore, we also notice a bias toward the Pro-Ukraine stance in our dataset. This is revealed by certain widely used words and specific emojis phrases with negative connotations toward the Russian invasion. We believe this general finding can support further studies on our dataset, e.g., misinformation detection for crisis events or social opinion analysis during the war.

\textbf{Limitations.} 
% Our dataset is limited by the loss of previously deleted posts and comments. We are not able to collect those posts and comments that have already been deleted before Pushshift scrapes them. The pushshift.io API collects data for users by performing queries in the Pushshift dataset \cite{baumgartner2020pushshift}. All collected data through this API actually has been scraped and stored in the Pushshift dataset previously. Therefore, only before the posts or comments are deleted can Pushshift scrapes and stores their data. Otherwise, the queries only return results showing \enquote{[deleted]} or \enquote{[removed]}. As a result, our dataset has also missed the contents of such deleted posts and comments for this limitation.
\james{Our dataset is limited by the loss of previously deleted posts and comments. Since our collection is based on pushshift.io API, we are not able to collect those posts and comments that have already been deleted before Pushshift scrapes them. In details, the API queries for such posts and comments only return results showing \enquote{[deleted]} or \enquote{[removed]}. As a result, our dataset can not collect or store these deleted contents.}

% \gareth{Can I suggest removing the above paragraph? It sounds like we are complaining about pushshift :) Instead, we could keep it simple and say that we miss posts that are deleted?}

In addition, our dataset does not include the two major Russia-specific subreddits, \emph{russia} and \emph{russiapolitics}. This is because Reddit has quarantined these communities due to the high volume of misinformation,\footnote{\url{https://www.reddit.com/r/russia/}} and Pushshift does not record any data from these subreddits.

\bibliographystyle{IEEEtran}
\bibliography{dataset}

\end{document}